\documentclass[a4paper,11pt]{article}
\pdfoutput=1

\usepackage{jheppub}
\usepackage{hyperref}
\usepackage[T1]{fontenc}
\usepackage{graphicx}
\usepackage{enumitem}
\usepackage{latexsym}
\usepackage{amsfonts}
\usepackage{amssymb}
\usepackage[dvipsnames]{xcolor}
\usepackage{amsmath}
\usepackage{slashed}
\usepackage{dcolumn}
\usepackage{verbatim}
\usepackage{float}
\usepackage{multirow}
\usepackage{xspace}
\usepackage[normalem]{ulem}
\usepackage[caption=false]{subfig}
\usepackage{bm}
\usepackage{bbm}

\newcommand{\beq}{\begin{equation}}
\newcommand{\eeq}{\end{equation}}
\newcommand{\bea}{\begin{eqnarray}}
\newcommand{\eea}{\end{eqnarray}}

 \def\d{{\rm d}}

\newcommand{\refjnl}[1]{{\rm#1}}

\def\apj{\refjnl{ApJ}}                 
\def\apjs{\refjnl{ApJS}}               

\usepackage{fancyvrb}

\RecustomVerbatimCommand{\VerbatimInput}{VerbatimInput}%
{fontsize=\footnotesize,
 frame=lines,  
 framesep=2em, 
 rulecolor=\color{Gray},
 labelposition=topline,
 commandchars=\|\(\), 
 commentchar=*        
}

\newcommand{\vtwo}[1]{{#1}}

\renewcommand{\paragraph}[1]{~\\ \noindent{\bf \emph{#1} --}}

\newcommand{\MPl}{M_{\rm Pl}}

\title{Repository for extended dark matter object constraints}
\author[a]{Djuna Croon,}
\author[a]{Sergio Sevillano Mu\~noz}

\affiliation[a]{Institute for Particle Physics Phenomenology, Department of Physics, Durham University, \\Durham DH1 3LE, U.K.
}

\emailAdd{djuna.l.croon@durham.ac.uk}
\emailAdd{sergio.sevillano-munoz@durham.ac.uk}
\preprint{IPPP/24/38}

\abstract{Extended dark matter objects (EDOs) are popular dark matter candidates that interact gravitationally with the Standard Model. These gravitational interactions can be used to constrain their allowed parameter space. However, EDOs can have different formation mechanisms, sizes, and shapes, requiring a case-by-case analysis when studying their impact on different areas of cosmology. We thus present a \href{https://github.com/SergioSevi/EDObounds}{repository} of all available bounds for these objects, with a code that allows plotting user-defined combinations of all up-to-date bounds for a given shape and different radii. We propose a standard for the EDOs' mass profiles so that different sets of bounds are consistent with each other, and provide instructions on using the code and contributing to the repository.

}
\begin{document}
\maketitle
\flushbottom

\section{Introduction}\label{section: introduction}
 While dark matter comprises most of the total amount of matter in the universe, it has not been directly detected other than through its gravitational interactions.
 Multiple candidates have been proposed with different mass ranges, having different indirect effects on our universe depending on their origin. In this work, we focus on dark matter structures in the high mass range, which do not add extra interactions into the Standard Model but still lead to strong gravitational imprints that can be observed, such as using gravitational (micro-)lensing techniques
(e.g. \cite{Alcock_1998,Niikura:2017zjd,Zumalacarregui:2017qqd,Niikura:2019kqi,Fairbairn:2017sil,Croon:2020ouk,Bai:2020jfm,Croon:2020wpr,CrispimRomao:2024nbr,DeRocco:2023hij}), or gravitational waves (e.g. \cite{Bird:2016dcv,Ali-Haimoud:2017rtz,Kavanagh:2018ggo,Bertone:2019irm,LIGOScientific:2019kan,Chen:2019irf,Franciolini:2021nvv,Croon:2022tmr}). 

These objects, commonly called MACHOs, are normally created in the early universe, and their most popular example are primordial black holes (PBHs)~\cite{Green:2020jor,Carr:2016drx}. However, we will focus on those massive compact objects that do not have a black hole in their interior, \vtwo{which we call} extended dark matter objects (EDOs). These are a popular dark matter candidate due to the different formation mechanisms they can have, which makes them very generic in principle, but has the downside of needing to be constrained case by case depending on their mass profile. Still, strong constraints have been placed using gravitational (micro-)lensing techniques~\cite{Croon:2020ouk,Croon:2020wpr}, gravitational waves~\cite{Croon:2022tmr}, accretion of baryonic matter onto the CMB spectrum~\cite{Bai:2020jfm,Croon:2024rmw} and dynamical heating of stars~\cite{Graham:2024hah}. \vtwo{Depending on the underlying theory producing the EDOs, they can be constrained by their introduction of beyond the standard model physics, as is the case for axion stars~\cite{Tkachev:2014dpa}, among others. However, here, we will only consider those constraints coming from their gravitational influence on our universe, which are guaranteed to be model-independent.}

Independent of the dark matter candidate, a large compound of different constraints is needed to cover their full parameter space. This creates an inconvenient aspect common to many dark matter candidates, as having all of the different sets of constraints at once is usually an onerous task, requiring one to have an updated track of all publications and to group all different bounds so they can be presented. Therefore, repositories are very beneficial for such communities, where a small individual effort makes it simple to have an updated record of all bounds for a given candidate. Among others, important examples include repositories for axion dark matter in Ref.~\cite{AxionLimits} and for PBHs in Ref.~\cite{Bradley_pbh}. Based on an extension of the latter example, we present the creation of a repository for EDOs. 

In this work, we first present the definition of the most well-constrained and popular EDO mass profiles in section~\ref{section: EDO}, which will be used as the standard in the repository. Then, in section~\ref{section: code}, we show how the repository works, both from a user and a contributor perspective, presenting the repository structure and some examples. Finally, we conclude in section~\ref{section: Conclusion}.

\section{Definition of EDO profiles}\label{section: EDO}
Macroscopic dark objects, with masses ranging from that of a small asteroid to that of a supermassive black hole, occur in many models of DM. While for point-like (and effectively point-like) objects such as PBHs the only relevant parameter is the mass,
for extended dark matter objects, we will need to constrain not only their mass and radii but also their shape. Indeed, an important aspect of EDOs is that the constraints may heavily depend on this shape of the mass profile, forcing us to present different sets of constraints in each case. This is a key difference with PBHs, and it is very important to define a standard so that all obtained constraints can be consistently plotted together. 

Each of these shapes differs depending on the formation mechanism for the object. So far, the most well-constrained mass profiles are the following:\footnote{We provide a Mathematica notebook with the definition of all these mass profiles in the \href{https://github.com/SergioSevi/EDObounds}{repository} for further clarity. }
\begin{itemize}
    \item \textbf{Navarro-Frenk-White (NFW) subhalos:} These profiles are mainly thought of as a density distribution that closely matches the cold dark matter clustering behaviour to explain {{(sub-)} structure} formation~\cite{NFW}. Here, we allow the mass and radius of these objects to vary considerably. The mass profile is given by 
    \begin{equation}
        M_{\rm NFW}(r)=\int_0^{r}\d \hat{r}\,4\pi\hat{r}^2 \rho_{\rm NFW}(\hat{r}),
        \label{eq:nfw}
    \end{equation}
    where
    \begin{equation}
        \rho_{\rm NFW}(\hat{r})=\frac{\rho_0}{\frac{\hat{r}}{R_s}\left(1+\frac{\hat{r}}{R_s}\right)^2},
    \end{equation}
    is the NFW density, with $R_s$ and $\rho_0$ parameters defining the total subhalo radius and mass via $R=100 R_s$ and $M=M_{\rm NFW}(100 R_s)$, respectively. Note that we choose to cut off the mass of this object at $100 R_s$, given that it technically diverges. This integral for the mass profile can be solved analytically, giving
    \begin{equation}\label{eq: NFW mass analytical}
        M_{\rm NFW}(r R_s)=4\pi \rho_0 R_s^3\left(-\frac{r}{1+r} + \log(1+r)\right).
    \end{equation}
     In particular, we are interested in $r_{90}$, defined as the radius that encloses the $90\%$ of the total EDO mass, and can be obtained by solving
    \begin{equation}
        \frac{\int_0^{r_{90}}\d \hat{r}\,4\pi\hat{r}^2 \rho_{\rm NFW}(\hat{r})}{\int_0^{100 R_s}\d \hat{r}\,4\pi\hat{r}^2 \rho_{\rm NFW}(\hat{r})}=0.9.
    \end{equation}
    Using the analytical solution from Eq.~\eqref{eq: NFW mass analytical}, we find that $r_{90}$ for the NFW subhalo is related to the EDO's total radius by $r_{90}\approx 0.69 R= 69 R_s$.
    \item \textbf{Boson stars:} Formed by the gravitational collapse of a scalar field~\cite{Schunck:2003kk}, boson stars are supported by the quantum mechanical effects inherent to bosons.
    Their mass profile can be obtained by solving Schrodinger's equation for a scalar field under the influence of its own gravitational potential~\cite{Bar:2018acw}. This is given by solving the following set of differential equations:
    \begin{align}
        i\partial_t \psi&= -\frac{1}{2m}\nabla^2 \psi +m \Phi \psi\nonumber\\
        \nabla^2 \Phi&= 4\pi G|\psi|^2, 
    \end{align}
    where $\psi$ is the field's wavefunction, $\Phi$ is the gravitational potential, $G$ is the Newton's gravitational constant and $m$ the field's mass as given by the Klein-Gordon equation, not to be confused with the EDO's total mass. This set of equations can be solved using the ansatz
    \begin{equation}
        \psi(x,t)=\left(\frac{m \MPl}{\sqrt{4 \pi}}e^{-i\gamma mt} \chi (x)\right),
    \end{equation}
    where $\MPl=1/\sqrt{G}$ and $\gamma$ is proportional to the EDO energy per unit mass. This allows to express the set of differential equations as
    \begin{align}
        \partial_r^2 (r\chi)&= 2r(\Phi-\gamma)\chi\nonumber\\
        \partial_r^2(r \Phi)&= r \chi^2. 
    \end{align}
    As the solution converges at infinity, we will define the EDO's total radius where it encloses $99.9\%$ of its mass. Similarly to NFW subhalos, we will constrain these objects for their $r_{90}$, for which a numerical calculation gives $r_{90}\approx0.53 R$.
\end{itemize}
Currently, these mass profiles are already well-constrained from the absence of micro-lensing~\cite{Croon:2020ouk,Croon:2020wpr}, gravitational wave production~\cite{Croon:2022tmr} and baryon accretion~\cite{Croon:2024rmw}. Additionally, bounds using the dynamic heating of stars were obtained only for the NFW subhalo profile~\cite{Graham:2024hah}. Other EDO mass-profiles that have been considered in some instances are:

\begin{itemize}
    \item \textbf{Uniform spheres (of constant density):} This is usually treated as a toy model due to its simplicity. However, this mass profile closely represents quark nuggets formed from standard~\cite{Witten:1984rs} or axion~\cite{Zhitnitsky:2002qa} QCD. It is defined as 
    \begin{equation}
    M_{\rm Uni}(r)=M \begin{cases}
        \left(\frac{r}{R}\right)^3 &\qquad r\leq R\\
        1 &\qquad R<r,
    \end{cases}
    \label{eq:uniformsphere}
    \end{equation}
    which has a constant density in its interior.

    \item \textbf{Ultra-compact mini-halos:} While large over-densities in the early universe usually lead to the creation of primordial black holes, there are cases in which the gravitational potential is not large enough to collapse into a singularity. In that case, for small enough over-densities, the relic would form a so-called ultra-compact mini-halo~\cite{Ricotti:2009bs}. The mass profile of these types of EDOs is well-described by~\cite{Bertshinger}
    \begin{equation}
    M_{\rm UCMH}(r)=M \begin{cases}
        \left(\frac{r}{R}\right)^{3/4} &\qquad r\leq R\\
        1 &\qquad R<r.
    \end{cases}
    \label{eq:ucmh}
    \end{equation}
\end{itemize}

For the last two cases, the mass function does not diverge at infinity, so there is no need to define an arbitrary point as an effective radius for the EDO. However, we will still classify these objects by their $r_{90}$, where they enclose $90\%$ of their mass. 
\begin{figure}
    \centering
    \includegraphics[width=0.5\textwidth]{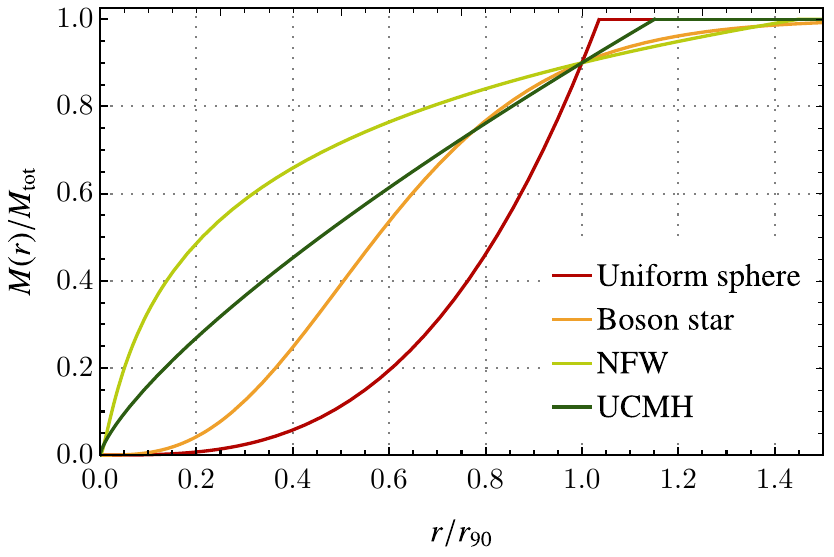}
    \caption{Mass profiles of the EDOs described in the text, as taken from Ref.~\cite{Croon:2024rmw}.}
    \label{fig:mass_profiles}
\end{figure}
We note that for present-day constraints, tidal stripping needs to be considered for the EDOs with the largest radii. Due to interactions with the host halo, these would not have survived and therefore cannot be constrained. The survival probability has specifically been calculated for NFW subhalos in \cite{Graham:2024hah,vandenBosch:2017ynq}. This primarily affects larger EDOs with radii $R > 10^6 \rm R_\odot$, though this may differ depending on both EDO mass and environment. 

Another common assumption for both EDOs and PBHs alike is that for sub-fractions of DM ($f_{\rm DM}<1$), the density of objects in a galaxy still follows the DM halo profile. Relaxing this assumption could, in principle, either weaken or strengthen the constraints on DM sub-fractions. \vtwo{Additionally, we will assume a monochromatic distribution of EDO masses for all cases. This approach may not represent the most realistic scenario in every context, but it avoids committing to any specific mass distribution, making it a useful middle ground for comparison. Moreover, as with PBHs, the bounds for other mass distribution functions can be derived from the monochromatic data~\cite{Carr:2017jsz}.}
    
In Figure~\ref{fig:mass_profiles}, we present the mass functions for all described objects. It is important to be precise in the EDOs' mass profiles when combining different sets of constraints. Thus, in what follows, we will explain how the repository of EDO bounds is classified and structured using the definitions presented in this section.

\section{Repository usage}\label{section: code}
Combining all available constraints into a repository is beneficial both to contributing authors, who can add and promote their obtained constraints, and to users, who can easily obtain and cite all existing bounds for specific objects. For this reason, we will use different subsections to describe the guidelines for each case. 

\subsection{User's guide} \label{subsection: code users}
From a user's perspective, the only important input is the so-called \textit{listfile}, a .txt file enclosing all the information about the EDO mass profile, radii (as given in $r_{90}$), and set of bounds one wants to plot together. We provide an example in the repository with all available sets of constraints, which has the following structure:

\begin{Verbatim}
#Shape: (Choose from NFW, Uniform, Boson or UCMH)
NFW

#Radius (r): rmin, rmax, rjumps, where R_90=10^r RSun
	   -1     6	1
       
# Bound,     linestyle,  x,         y,        rotation, display name
CMBAccretion.   -       700	 1e-3	-85      CMB
EROS-2	  --      1e-2        1e-2	0	EROS-2
OGLE-IV	 --      1e-5        7e-3	0	OGLE-IV
Subaru-HSC.     --      1e-9	3e-3	0	Subaru-HSC
Lensing         -       1e-5	7e-3	0	Lensing
Dynamic-heating --       50	 5e-2	-65      Dynamic_Heating
LVK             -       1e-5        7e-3        0        LVK
\end{Verbatim}
\begin{figure}
    \centering
    \includegraphics[scale=0.6]{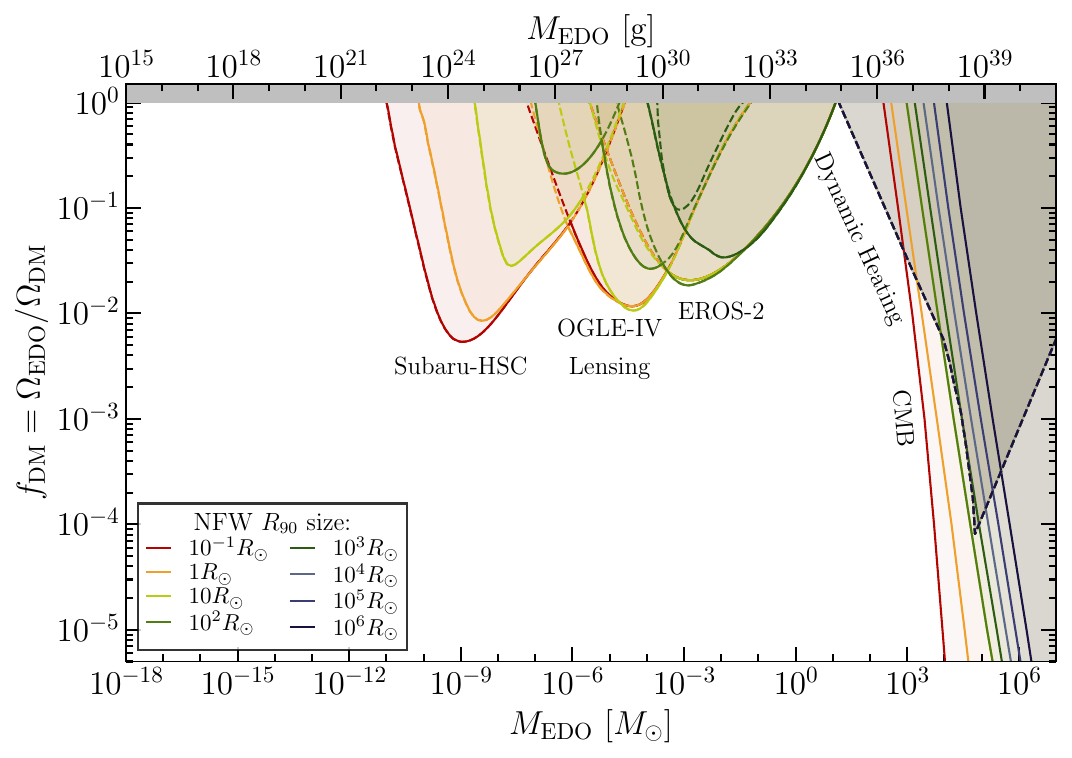}
    \caption{Example of a plot produced using our repository code with the listfile \texttt{bounds\_all.txt}, containing all current bounds on EDOs~\cite{Croon:2024rmw,Croon:2020wpr,Croon:2020ouk,Graham:2024hah} at these scales (since gravitational waves constrain more compact objects, the code automatically removed the label). The list of citations was automatically generated by the repository in the \texttt{Cite.txt} file, as mentioned in the text. Moreover, we used the \texttt{Lensing} bound to get the minimum of all micro-lensing constraints. We used dashed lines for dynamic accretion~\cite{Graham:2024hah} as these were calculated for a different radius cut-off for the NFW shape (this being $R_s$ instead of $100R_s$ as described in the previous section).
    }
    \label{fig: boundsall}
\end{figure}
This repository is an extension of the PBH constraints repository created by Bradley Kavanagh~\cite{Bradley_pbh}, so readers familiar with that repository will find most aspects to be similar. The main difference is that, contrary to PBHs, we need to specify the object's mass profile and the set of radii we want to plot. As we can see, the first line allows the user to choose the mass profile using one of the different shapes defined in the previous section. Then, the next entry specifies the different displayed radii, where we need to input $r_{90}$ in powers of 10 Solar radii. The three numbers in the entry correspond to the lowest radius, highest radius and the difference between each plotted radii, respectively. Finally, the last part sets the constraints we want to add to the plot, where different columns change the bounds display as follows: \texttt{Bound} corresponds to the name as it appears in the files folder;\footnote{Using the \texttt{Lensing} bound shows all lensing bounds, currently being \texttt{EROS-2}, \texttt{OGLE-IV} and \texttt{Subaru-HSC}, and gives the combination of these as a single constraint. \vtwo{ Similarly, using \texttt{All} will combine all bounds as a single constraint.}} \texttt{linestyle} allows the change of the plotting line style; \texttt{x}, \texttt{y}, and \texttt{rotation} set the position and inclination of the displayed text, defined by the last column. Contrary to PBHs bounds, we do not allow to change the colour of a bound since this degree of freedom will be used to plot multiple radii at once.

Once a listfile is created, the plot containing all constraints is obtained by running in the terminal
\begin{verbatim}
  python PlotEDObounds.py --listfile LIST_FILE --outfile OUT_FILE
\end{verbatim}
where \texttt{LIST\_FILE} is the .txt file defined previously and \texttt{OUT\_FILE} is the name for the generated plot. Additionally, this code automatically creates the list of \textit{bibitems} for all bounds appearing in your plot, which can be found in the file called \texttt{Cite.txt}. The produced plot using the file \texttt{bounds\_all.txt} is shown in Figure~\ref{fig: boundsall}.
  
\subsection{Contributing to the repository}\label{subsection: code contribute}

Contributions to the repository can be made by generating the necessary folder with all available bounds that apply to any specific shape and radii. So that it is easier to add into the system, we structure the files as \texttt{Bound\_name}>\texttt{Shape}>\texttt{radii}, as can be seen in the example from Figure~\ref{fig: bounds folder}. 

Setting each folder's and file's names correctly is very important. In particular, the bound name will be used to include the bounds in the listfile, so giving them a descriptive name is key. Inside this folder, we first classify all the bounds depending on their mass profiles, using the name codes of \texttt{NFW}, \texttt{Boson}, \texttt{Uniform} or \texttt{UCMH}. Additionally, at that level, we recommend adding a text file named \texttt{Citation.txt}, which has a small description of the set of bounds and the \textit{bibitem} of your paper, so that it will be added to the \texttt{Cite.txt} file every time a user includes your bounds when making a plot. As an example, this is how it looks for the CMB accretion~\cite{Croon:2024rmw} bounds:
\begin{figure}
    \centering
    \boxed{\includegraphics[width=0.98\linewidth]{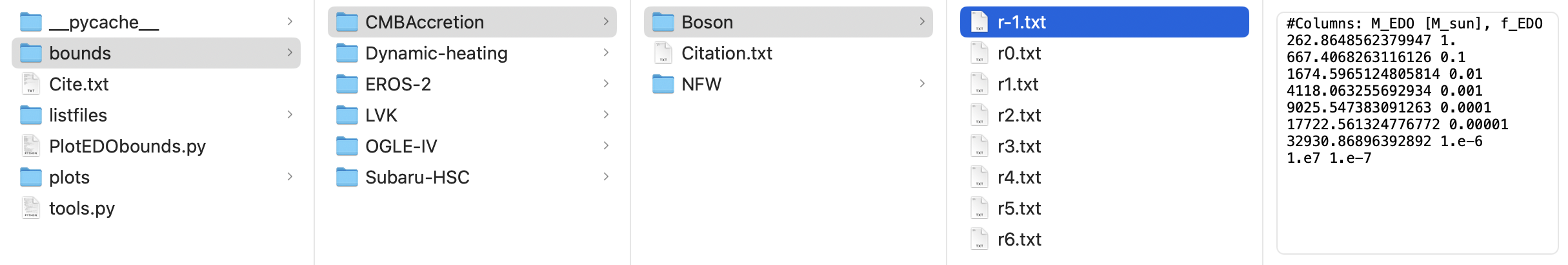}}
    \caption{Structure of the files inside the EDO repository. So that it is easy to incorporate new bounds, we have different shapes and radii options inside the folder with the name corresponding to the set of bounds. }
    \label{fig: bounds folder}
\end{figure}
\begin{Verbatim}
#Fig.7 in ArXiv:2403.13072
#CMB bounds due to accretion of baryonic mater around EDOs, altering recombination history.

@article{Croon:2024rmw,
    author = "Croon, Djuna and Sevillano Mu\~noz, Sergio",
    title = "{Cosmic microwave background constraints on extended dark matter objects}",
    eprint = "2403.13072",
    archivePrefix = "arXiv",
    primaryClass = "astro-ph.CO",
    reportNumber = "IPPP/24/11",
    month = "3",
    year = "2024"
}
\end{Verbatim}

\begin{figure}
    \centering
    \includegraphics[scale=0.4]{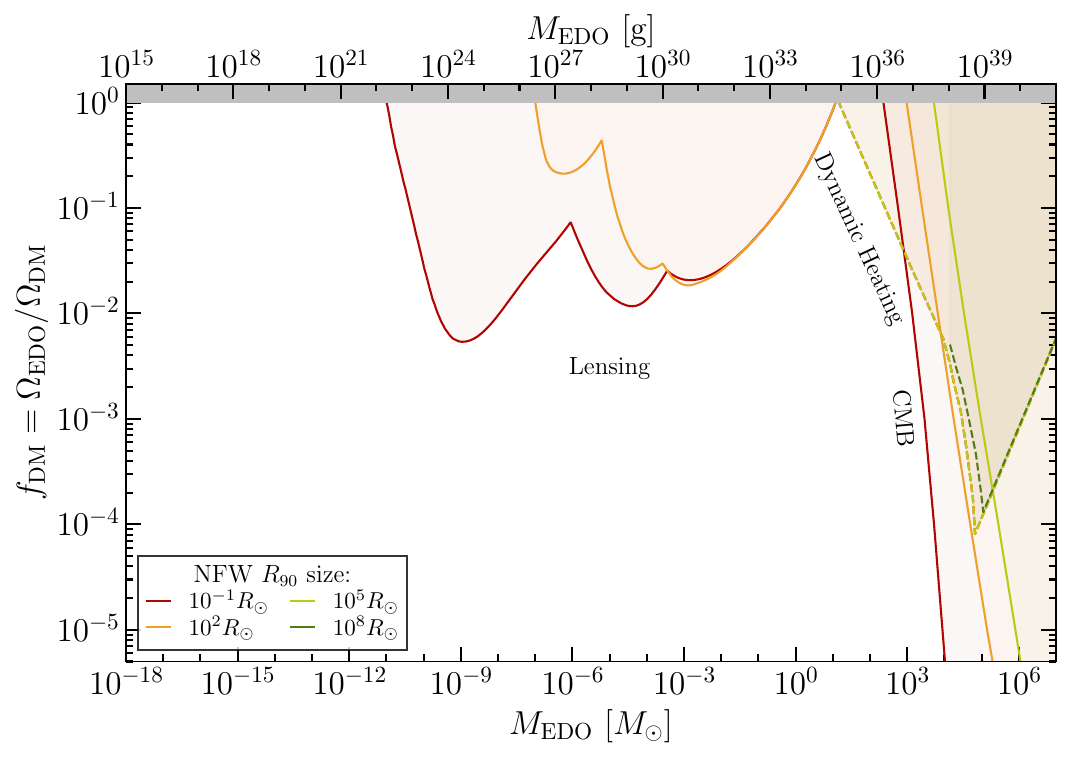} \includegraphics[scale=0.4]{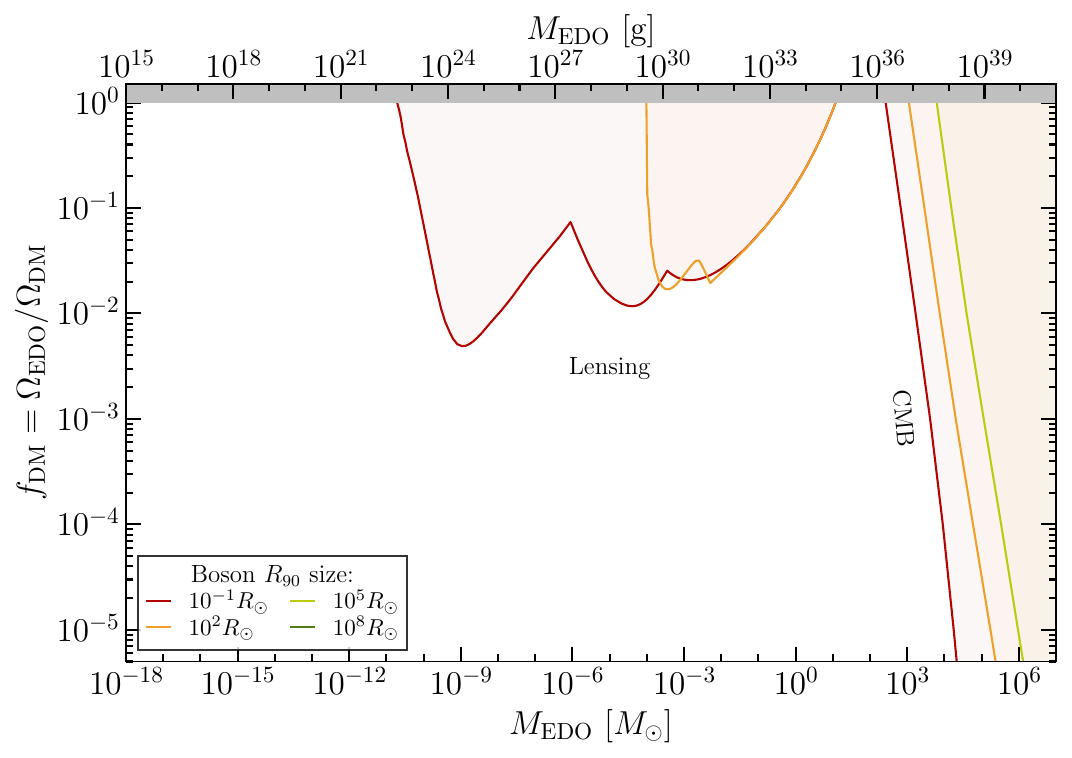}
    \caption{Example for generated plots with an incomplete set of bounds. On the left, we have an NFW sub-halo, where it is clear that the Lensing and CMB bounds do not cover the whole range of plotted radii due to the nature of constraints. On the right, in addition to this lack of constraints on Lensing and CMB, there are no bounds on dynamic heating as they have not yet been calculated for these objects.}
    \label{fig:example incomplete plots}
\end{figure}

Inside each of the mass profile folders, there must be one \texttt{.txt} file containing the set of bounds for each radius. To make it easy to track, the name of each file must be \texttt{rN.txt}, where \texttt{N} is the power of 10 for $r_{90}$ in Solar radius ($\texttt{N}=\log_{10}(r_{90}/R_\odot)$), where negative numbers have a hyphen in front (e.g., \texttt{r-1.txt} for $r_{90}=10^{-1}R_\odot$). Inside each of these files, the data is displayed in two columns, left and right, corresponding to $M_{\rm EDO}$ (in Solar masses) and {the dark matter fraction corresponding to EDOs} ($f_{\rm DM}$), respectively. As an example, this is the \texttt{r-1.txt} file for \texttt{NFW} shape in \texttt{CMBAccretion}:
\begin{Verbatim}
#Columns: M_EDO [M_sun], f_EDO
213.350665152798 1.
529.8498935703813 0.1
1269.3708748549936 0.01
2884.035135358922 0.001
5777.9293930551885 0.0001
9611.252530758793 0.00001
13853.125643302837 1.e-6
1.e7 1.e-7
\end{Verbatim}

Once a complete bounds folder is created, the instructions to add it to the repository can be found on \href{https://github.com/SergioSevi/EDObounds}{github.com/SergioSevi/EDObounds}. Having bounds for all different radii and aforementioned mass profiles is not necessary to contribute to this repository. As long as the bounds correspond to one of the mass profiles described earlier, they will automatically add to the plot to which they correspond\footnote{\vtwo{In the case these bounds cannot be extended to the mass profiles described in Section~\ref{section: EDO}, a new folder can be added to the repository for a different EDO shape, if necessary. However, we believe it is advantageous to generate all constraints for the same type of object, allowing for easier comparison across bounds.}}. For example, in Figure~\ref{fig:example incomplete plots}, we can see that some bounds only exist for one of the mass profiles and others only for the smallest sizes. Additionally, if the set of constraints converges beyond a certain critical radius, the code will automatically apply these bounds to any smaller radii. 

\section{Conclusion}\label{section: Conclusion}
Dark matter candidates interacting solely gravitationally with the Standard Model leave many different imprints in our Universe that can be used to constrain them. It is thus difficult to maintain an up-to-date compilation of bounds for the same candidate. For this purpose, it is common to create repositories that allow plotting all bounds for a certain dark matter candidate, such as those for axions~\cite{AxionLimits} or PBHs~\cite{Bradley_pbh}. In this paper, we have presented the creation of a repository for extended dark matter objects.

Extended dark matter objects are a subset of MACHOs which do not behave as point-like masses. They are usually created in the early universe through a variety of mechanisms, which lead to different mass profiles associated with them. Here, we aim to create a standard of the best-motivated mass profiles so all obtained constraints can be applied to the same type of objects. In particular, we have focused on NFW subhalos and boson stars, but we have also included those EDOs with constant density or ultra-compact mini-halos. A big difference with PBHs is that these objects can have different radii, requiring different constraints for each. So far, different bounds have been found for these objects, such as via the absence of gravitational waves from binary mergers~\cite{Croon:2022tmr}, (micro-)lensing~\cite{Croon:2020ouk,Croon:2020wpr}, CMB accretion~\cite{Bai:2020jfm,Croon:2024rmw}, and dynamical heating of stars~\cite{Graham:2024hah}. 

In this paper, we have also provided instructions for using our repository, which is a direct extension of that by Bradley Kavanagh~\cite{Bradley_pbh}, but includes all necessary additions to allow for plotting different EDO radii simultaneously. Moreover, we included the generation of a bibliography in each generated plot so that all bounds are accordingly cited. The instructions to contribute to the repository, and the repository itself, can be found at \href{https://github.com/SergioSevi/EDObounds}{github.com/SergioSevi/EDObounds}.
\section*{Acknowledgments}
We thank Bradley Kavanagh and Harikrishnan Ramani for useful discussions. DC and SSM are supported by the STFC under Grant No.~ST/T001011/1.
\bibliographystyle{apsrev4-1}
\bibliography{main}
\end{document}